# Direct observation of the enhanced photonic spin Hall effect in a subwavelength grating supporting surface plasmon resonance

N. I. Petrov [1], Y. M. Sokolov [1], V. V. Stoiakin[1], V. A. Danilov[1], V. V. Popov[2] and B. A. Usievich[3]

[1] Scientific and Technological Centre of Unique Instrumentation of the Russian Academy of Sciences, Moscow 117342, Russia

[2]Lomonosov Moscow State University, Moscow 119991 , Russia

[3]Prokhorov General Physics Institute of the Russian Academy of Sciences, Moscow 119991, Russia

**Abstract**: The photonic spin Hall effect (PSHE) in surface plasmon resonance (SPR) structures has great potential for various polarization-sensitive applications and devices. Here, using optical weak measurement, we observe spin-dependent and spin-independent angular shifts of the reflected beam, enhanced by SPR in a subwavelength nickel grating. An enhanced in-plane photonic spin Hall effect manifested in the angular splitting of circularly polarized photons with opposite helicity signs is demonstrated. We theoretically and experimentally demonstrate that angular in-plane shifts can be changed from spin-independent (Goos-Hänchen shift) to spin-dependent (PSHE) when the incident beam polarization state changes. The SPR-induced depolarization of light and the mixing of polarization states are detected. High purity of spin separation and a high degree of circular polarization are achieved with an optimal polarization state (preselection angle) and a resonance angle of incidence. The spineless spatial separation of two orthogonal components of the field with diagonal linear polarizations is demonstrated.

## 1. Introduction

A light beam being reflected or refracted at the interface, experiences spatial and angular shifts depending on its polarization and intensity profile. Both reflected and refracted beams are formed as a result of interference of waves reradiated in neighboring media. The analysis revealed many interesting phenomena, including the well-known spin-independent Goos-Hänchen (GH) and spin-dependent Imbert-Fedorov (IF) shifts in the plane and out-of-plane of propagation of the incident beam, respectively [1-4]. A significant increase in the GH shift is achieved by excitation of surface plasmon waves [5-7] and Bloch surface electromagnetic waves in photonic crystals [8]. A noticeable increase in the angular displacement can be observed for a tightly focused beam reflected by a dielectric surface near the Brewster angle [9]. The physical origin of the lateral GH shift lies in the angular dispersion of the Fresnel reflection coefficients. The IF shift is a spin-dependent effect that occurs as a result of spin-orbit interaction [4]. In the last decade, the spin Hall effect of light has been intensively studied in various physical systems [10-35].

Usually, PSHE as well as IF shift is associated with the spatial and angular separation of photons with opposite spin angular momentum (SAM). Unlike the IF shift, the PSHE shift can be either transverse or longitudinal. Spin-dependent transverse shifts between the left- and right-handed circularly polarized components were observed for photons passing through the air-glass interface [13-15]. Subsequently, the longitudinal (in-plane) photonic spin splitting was observed [16-19]. The large transverse PSHE was demonstrated for a beam passing through a tilted polarizer [20, 21]. A significant enhancement of the PSHE occurs via the surface plasmon resonance (SPR) [22-32]. A 0.16 mm displacement of separated reflected light spots corresponding to opposite circular polarizations was observed in a planar waveguide via ultrahigh-order modes [35]. Recently, the PSHE in an anisotropic metamaterial has been considered as an effective way for image edge detection [36-38]. The PSHE has also been considered in a dielectric grating and in compound grating waveguide structures [39, 40]. In [40], a polarization-dependent bound state in the continuum (BIC) in a grating waveguide structure is theoretically proposed. It has been shown that the transverse shift of the PSHE, due to the unique resonance property and polarization-dependent property of the quasi-BIC, can be intensively enhanced to the order of hundreds of micrometers.

Of particular interest is the search for effective ways to control the spin angular shift, especially to increase its magnitude or switch the spatial position of the reflected beam. Currently, several methods of controlling the spin angular shift have been theoretically proposed, including changing the complex refractive index [28], the incident beam waist and dielectric constant of the metamaterial [33], and controlling the Brewster angle [34]. However, these methods have not yet been experimentally tested on subwavelength gratings supporting SPR, taking into account the influence of the parameters of the diffraction grating (depth, period, dielectric constant of the material) and the incident beam (wavelength, waist size, angle of incidence).

In this paper, we demonstrate the spin Hall effect of light in the light reflection from a subwavelength nickel grating enhanced by SPR. Using the standard PSHE theory, we show that in-plane angular shifts can be changed from spin-independent (Goos-Hänchen shift) to spin-dependent (PSHE) when the polarization state of the incident beam slightly changes. Unlike the spin-dependent out-of-plane IF shift, here we investigate spin-dependent in-plane shift enhanced by SPR in subwavelength grating. The effects of angular shift and beam shape change for a reflected beam near surface plasmon resonance conditions are demonstrated experimentally and confirmed by rigorous electromagnetic simulations. We verify our simulations experimentally using the polarimetric and quantum weak measurement methods. SPR-induced depolarization of light and mixing of polarization states are detected. A novel phenomenon of the spin-independent splitting of two orthogonal field components with linear polarizations is described.

## 2. Experimental setup and theory

In Fig. 1a the experimental setup consisting of a He-Ne laser, half-wave and quarter-wave plates, a linear polarizer, a focusing lens, a subwavelength diffraction grating, and a CCD camera for measuring the beam profile is presented. The angular resolution of the motorized rotating platform (ZX110-100, China) on which the diffraction grating is mounted is 0.0005°. The angle measurement error (repeatability) is 0.005°=18". Fig. 1b illustrates the reflection of the beam from the air-grating interface. The coordinate frames ($x_i$, $y_i$, $z_i$) and ($x_r$, $y_r$, $z_r$) correspond to the incident and reflected electric fields.

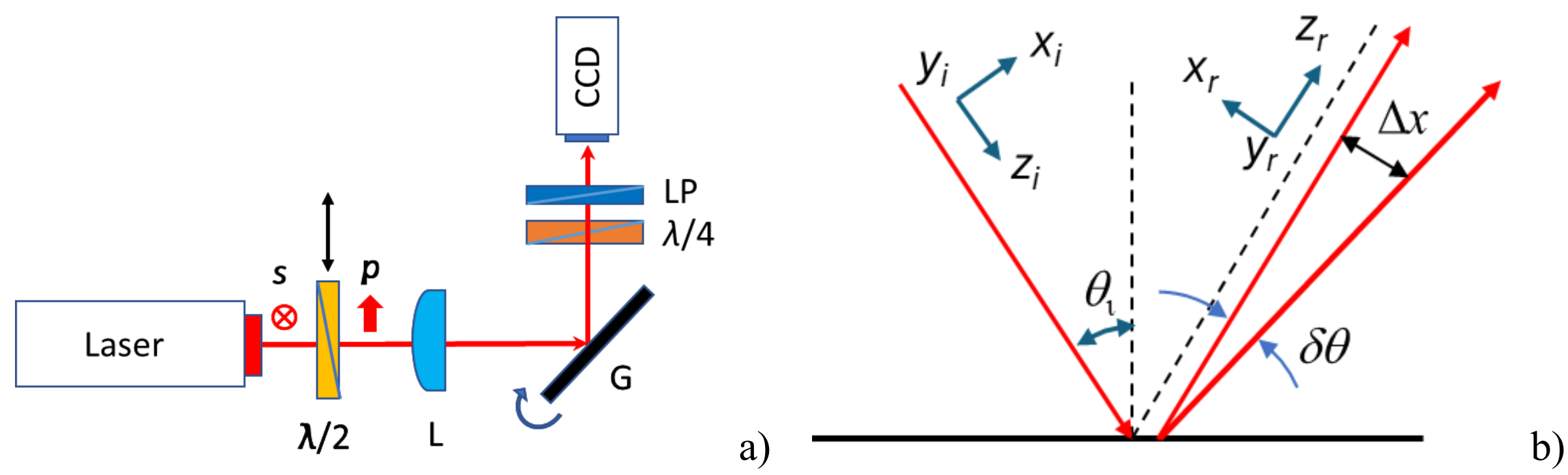

**Fig. 1.** (a) Experimental setup. $\lambda/2$ – half-wave plate, L – lens, G – subwavelength grating, $\lambda/4$ – quarter-wave plate, LP – linear polarizer, and CCD – camera. (b) Schematic diagram of the in-plane PSHE when the beam is reflected from the grating surface. The $y_i$ and $y_r$ axes are directed perpendicular to the plane of incidence.

We investigate angular shifts when the surface plasmon resonance is excited in a subwavelength nickel grating, which is a 1D all-nickel substrate with a periodically corrugated surface (Fig. 2). Note that the angular shift is easier to measure because of an additional enhancement caused by the propagation factor. The grating was made by using electron beam lithography (e-beam) and a layer of negative photoresist (polymethyl-methacrylate, PMMA) deposited on a nickel substrate using a spin-coater [41]. Then, the photoresist was etched to form one-dimensional grating patterns. The sample and its spatial profile were characterized by scanning electron microscopy (SEM) and atomic force microscopy (AFM). The AFM image of the subwavelength grating is shown in Fig. 2. The AFM cross-section has a close sinusoidal profile with a filling ratio of 0.5.

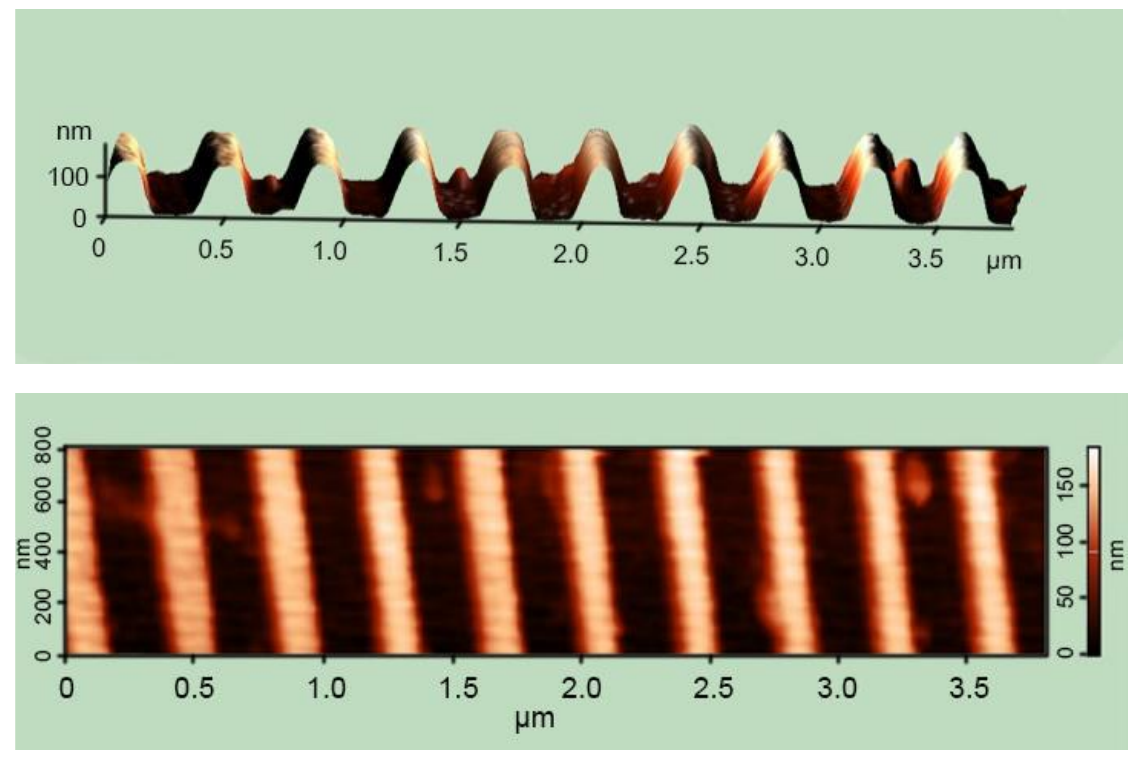

**Fig. 2.** AFM image of a subwavelength grating with a period Λ = 400 nm.

The electric field of the reflected beam can be described using the standard SHEL theory [4, 13, 16]. The complex amplitude of the reflected beam can be expressed as Fourier transformation:

$$\boldsymbol{E}(x_r, y_r, z_r) = \iint \boldsymbol{E}(k_{rx}, k_{ry}, z_r) e^{ik_{rx}x + ik_{ry}y} dk_{rx} dk_{ry}, \tag{1}$$

where the angular spectrum in the paraxial approximation is given by

$$\boldsymbol{E}(k_{rx}, k_{ry}, z_r) = \boldsymbol{E}_0(k_{rx}, k_{ry}) e^{iz_r\sqrt{k_0^2 - k_{rx}^2 - k_{ry}^2}} \cong \boldsymbol{E}_0(k_{rx}, k_{ry}) e^{ik_0 z_r} e^{-\frac{i}{2k_0}(k_{rx}^2 + k_{ry}^2) z_r}, \tag{2}$$

where $k_0 = 2\pi/\lambda$ is the incident wave vector, and $k_{rx}$, $k_{ry}$ are the components of the wave-vector of the reflected beam in $x_r$, $y_r$ direction.

Based on the boundary condition, the reflected angular spectrum is expressed as [13, 16]:

$$\begin{bmatrix} E_r^p \\ E_r^s \end{bmatrix} = \begin{pmatrix} r_p & \frac{k_{ry}(r_p + r_s)\cot\theta_i}{k_0} \\ -\frac{k_{ry}(r_p + r_s)\cot\theta_i}{k_0} & r_s \end{pmatrix} \begin{bmatrix} E_i^p \\ E_i^s \end{bmatrix}, \tag{3}$$

where $r_p$ and $r_s$ are the Fresnel reflection coefficients for $p$- and $s$- polarizations, respectively, $\theta_i$ is the incident angle, $k_{rx} = -k_{ix}$, and $k_{ry} = k_{iy}$.

Decomposing into a Taylor series in the vicinity of the central wave vector $k_{ix0} = k_0 \sin\theta_0$, $r_p$ and $r_s$ can be represented as:

$$r_{p,s}(\theta_i) = r_{p,s}(\theta_0) + (\theta_i - \theta_0) \left.\frac{dr_{p,s}}{d\theta_i}\right|_{\theta_i = \theta_0} + \frac{1}{2}(\theta_i - \theta_0)^2 \left.\frac{d^2 r_{p,s}}{d\theta_i^2}\right|_{\theta_i = \theta_0} \cdots . \tag{4}$$

Consider the incident $p$- polarized beam with a Gaussian angular spectrum:

$$E_i^p = \frac{w_0}{\sqrt{2\pi}} \exp\left[-\frac{w_0^2(k_{ix}^2 + k_{iy}^2)}{4}\right], \tag{5}$$

where $w_0$ is the beam waist.

Calculating the integral in Eq. (1) using Eqs. (3) - (5) and relationships $E_r^p = \frac{1}{\sqrt{2}}(E_r^+ + E_r^-)$ and $E_r^s = \frac{i}{\sqrt{2}}(E_r^- - E_r^+)$, we obtain the expression for the right- and left- handed circularly polarized components of the reflected field

$$E_r^{\pm} = \frac{1}{\sqrt{2\pi} w_0} \frac{z_R}{z_R+iz_r} \exp\left(-\frac{k_0}{2}\frac{x_r^2+y_r^2}{z_R+iz_r}\right)\left[r_{p,s} - \frac{ix_r}{z_R+iz_r}\frac{dr_{p,s}}{d\theta_i} \mp \frac{y_r \cot\theta_i}{z_R+iz_r}\left(r_p + r_s\right)\right], \quad (6)$$

where the positive and negative signs denote the right circularly polarized (RCP) and the left circularly polarized (LCP) field components, and $z_R = k_0 w_0^2/2$ is the Rayleigh length.

A rigorous electromagnetic theory based on the *C*- method is used for the calculations of the reflection coefficients $r_p$ and $r_s$ [5, 6]. In Fig. 3 the calculated and measured dependences of the reflectance (diffraction efficiency of zero-order) depending on the angle of incidence for a nickel grating with a period of $\Lambda$ = 400 nm and a depth of $h$ = 80 nm at a wavelength of radiation $\lambda$ = 632.8 nm with *p*- polarization are presented. The dielectric constant of nickel at the wavelength $\lambda$ = 632.8 nm is $\varepsilon$ = -10.1 + i14.73. Note that an *s*- wave does not excite surface plasmon waves, so a *p*- polarized incident wave is necessary to observe the surface plasmon resonance.

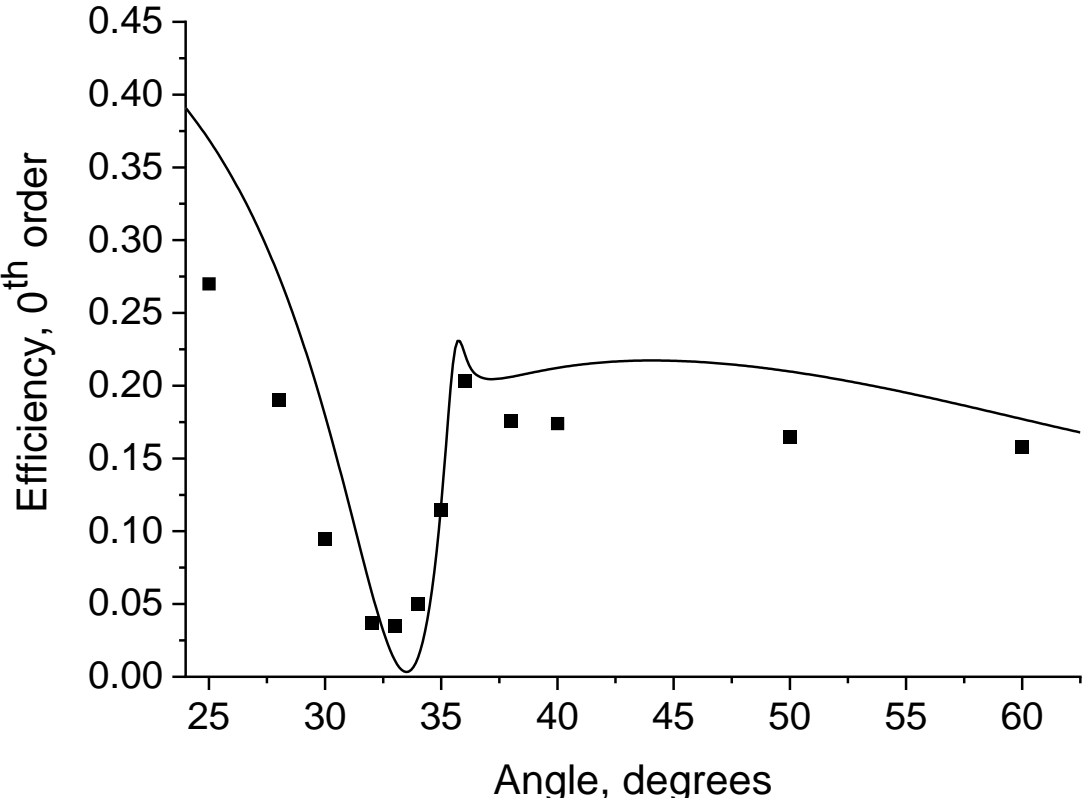

**Fig. 3.** Calculated and measured diffraction efficiencies of zero-order depending on the angle of incidence for a nickel grating with a period of $\Lambda$ = 400 nm and a depth of $h$ = 80 nm at a wavelength of radiation $\lambda$ = 632.8 nm with *p*- polarization.

Optimal conditions for the coupling of incident light with a surface plasmon wave is achieved at a grating depth of $h$ = 80 nm and a resonant angle of incidence of $\theta_{res}$ = 33.53° (Fig. 3), corresponding to the optimal condition for the coupling of incident light with a surface plasmon. In this case the minimum reflection from the grating is achieved. As the grating depth differs from the optimal, the value of the minimum reflection coefficient increases. The optimal depth depends on the grating period and material [5, 6]. It was shown that the optimal depth for the silver grating is $h$ = 20 nm.

In Fig. 4 the dependences of Fresnel reflection coefficients on the angle of incidence are presented.

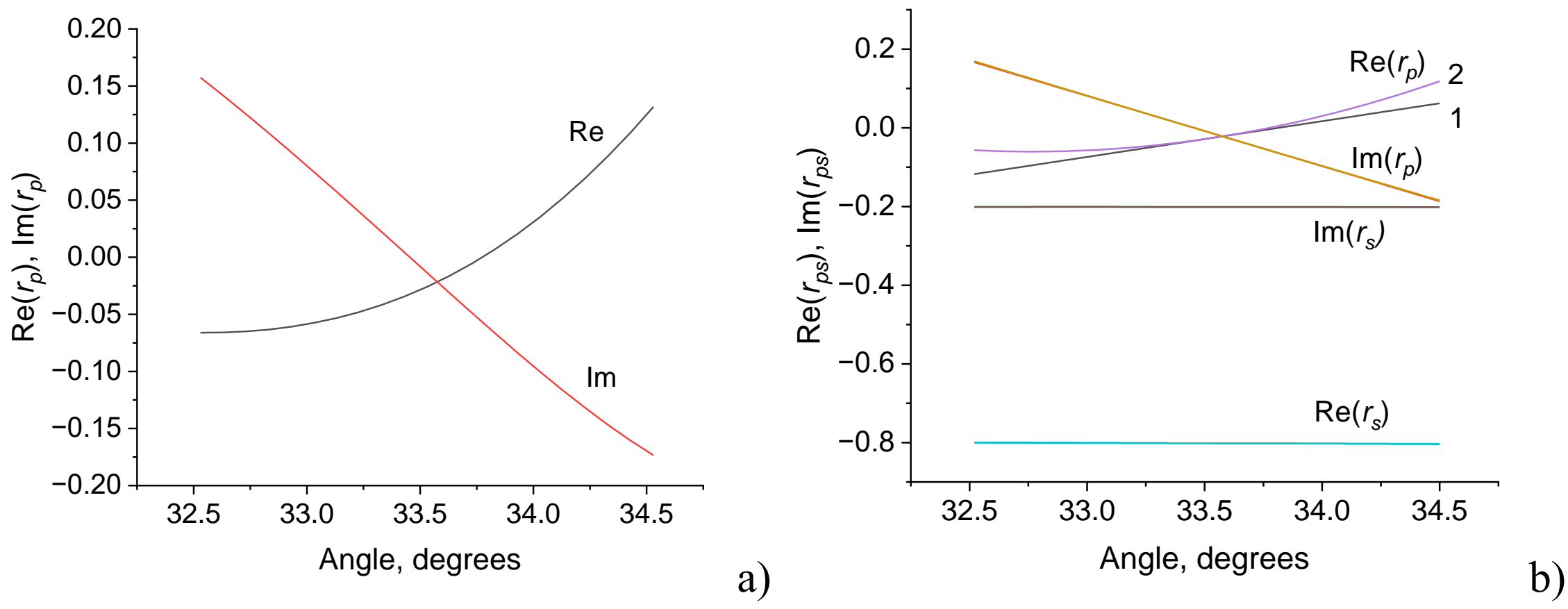

**Fig. 4.** The Fresnel reflection coefficients as function of incidence angle. (a) C-method, (b) using the expression (4). 1 – first-order approximation, 2 – second-order approximation.

Note that the coefficients Re($r_s$) and Im($r_s$), calculated using the rigorous C-method and expression (4), are practically the same. However, the reflection coefficients Re($r_p$) and Im($r_p$), calculated using the rigorous C-method, coincide with the approximate values (4) only near the central wave vector. Curve 2 in Fig. 4b is constructed taking into account the second-order term of smallness in (4). This indicates that for large deviations of the angle of incidence, Fresnel reflection coefficients obtained using rigorous calculations should be used. In this case, the reflected field can only be obtained by numerically calculating the integral (1).

## 3. Results and discussion

In Fig. 5, the measured and calculated intensity profiles of left- handed circularly polarized component of reflected beams passing through the left-handed circular polarization (LCP) filter, which consists of a quarter-wave plate and an adjusted linear polarizer, are presented at a distance of $z$ = 25 cm from the grating surface for different angles of incidence. An incident Gaussian beam with a beam waist $w_0$ = 22.6 $\mu$m is $p$- polarized. Two equal peaks appear when the angle of incidence is exactly equal to the resonant angle (Fig. 5b). The appearance of two spots around the resonant angle of incidence is due to the presence of a dip in the reflectivity curve (Fig. 3). While the central part of the dip with a minimum reflection coefficient leads to the almost disappearance of the reflected light from the incident Gaussian beam, the left and right sides of the dip with a higher reflectivity cause to the formation of two peaks. It follows from the measurement results that two spots with equal intensity are also observed after the passage of the reflected beam through the right-handed circular polarization filter (RCP filter). This indicates that in-plane spin-independent beam splitting (without separation of photons with the opposite SAM), which is a GH shift, occurs for an incident $p$-polarized beam at a resonant angle of incidence. It is seen from Figs.

5a and 5c, that the deviation of the angle of incidence in one direction or another from the resonant angle leads to a change in the spatial position of the reflected beam.

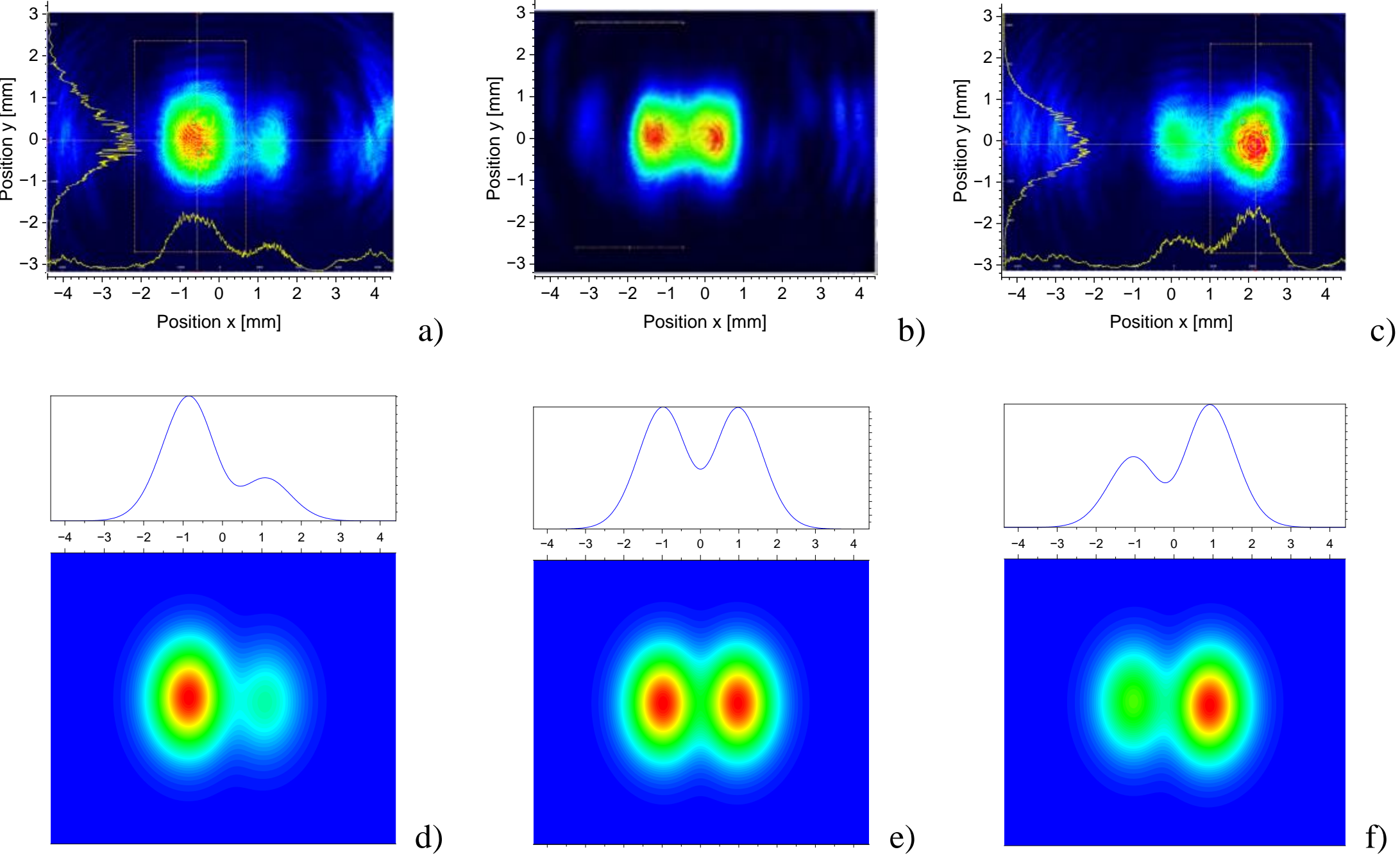

**Fig. 5.** Measured (a, b, c) and calculated (d, e, f) intensity profiles of reflected beams passing through the LCP filter. (a, d) $\theta_i = 33.48°$, (b, e) $\theta_i = 33.53°$, (c, f) $\theta_i = 33.58°$.

When the angle of incidence changes even by 0.01°, a significant change in the profile of the reflected beam occurs. This can be useful in the development of optical switches, as well as in angular metrology. To observe the enhanced in-plane spin-dependent splitting, an optimal overlap of the preselection and post-selection states should be created. This makes it possible to observe the enhancement of spin Hall shift by controlling the polarization of the incident beam, that is, by slightly changing the direction of the polarization vector of a linearly polarized incident beam relative to *p*-polarization. In this case, the incident beam becomes elliptically polarized (contains *p*- and *s*- polarization components) and the reflected beam is expressed as:

$$\begin{bmatrix} E_{rw}^{p} \\ E_{rw}^{s} \end{bmatrix} = \begin{bmatrix} r_p \cos\gamma + \frac{k_{ry}(r_p + r_s)\cot\theta_i \sin\gamma}{k_0} \\ r_s \sin\gamma - \frac{k_{ry}(r_p + r_s)\cot\theta_i \cos\gamma}{k_0} \end{bmatrix} \tilde{E}_i(k_{ix}, k_{iy}), \quad (7)$$

where $\gamma$ is the linear polarization angle of incident beam (preselection angle) between the polarization vector and axis $x_i$, and $\tilde{E}_i = \frac{w_0}{\sqrt{2\pi}} \exp\left[-\frac{w_0^2\left(k_{ix}^2 + k_{iy}^2\right)}{4}\right]$.

For this preselection, the in-plane spin splitting undergoes significant amplification. It follows from Eq. (7) that the center of gravity (centroid position) of the circularly polarized component of the reflected beam is expressed as

$$\langle x^{\pm}\rangle = \frac{\langle E^{\pm}|x|E^{\pm}\rangle}{\langle E^{\pm}|E^{\pm}\rangle} = \frac{2\sqrt{z_R^2+z_r^2}\,|\dot{r}_p|\cos\gamma(a_p\cos\gamma\mp a_s\sin\gamma)}{2k_0 z_R(a\pm b)+|\dot{r}_p|^2\cos^2\gamma}, \tag{8}$$

where $E^{\pm}$ are the electric fields that are right- and left- handed circularly polarized, respectively, $a_p = -|r_p|\sin(\varphi_p + \gamma_1 - \gamma_2)$; $a_s = |r_s|\cos(\varphi_s + \gamma_1 - \gamma_2)$; $a = |r_p|^2\cos^2\gamma + |r_s|^2\sin^2\gamma$; $b = |r_p||r_s|\sin(2\gamma)\sin(\varphi_p - \varphi_s)$, $z_R + iz_r = \sqrt{z_R^2 + z_r^2}\,e^{i\gamma_1}$, $z_R = k_0 w_0^2/2$, $r_p = |r_p|e^{i\varphi_p}, r_s = |r_s|e^{i\varphi_s}$, $\varphi_s$ and $\varphi_p$ are the phases of the Fresnel reflection coefficients $r_p$ and $r_s$, respectively, $\dot{r}_p = |\dot{r}_p|e^{i\gamma_2}$, $\dot{r}_p = \frac{\partial r_p}{\partial \theta_i}$.

At the optimal preselection angle, we clearly observe the in-plane angular splitting of reflected light with opposite circular polarizations. The optimal preselection angle corresponds to the condition under which the amplitude of reflected light with *p*- polarization becomes equal to the amplitude of reflected light with *s*-polarization.

In Fig. 6, the in-plane spin splitting for the preselection angle $\gamma = 4°$ is shown. When the reflected beam passes through a filter of a certain circular polarization, one of the spots disappears. This means that there is a spatial separation of left-handed and right-handed circularly polarized photons. At a preselection angle of $\gamma = 4°$, the right spot is left-handed circularly polarized, while the left spot is right-handed circularly polarized. On the contrary, at $\gamma = -4°$, the right spot is right-handed circularly polarized, and the left spot is left-handed circularly polarized. The physical mechanism of such separation is related to the spin-orbit interaction enhanced by surface plasmon resonance in a subwavelength diffraction grating.

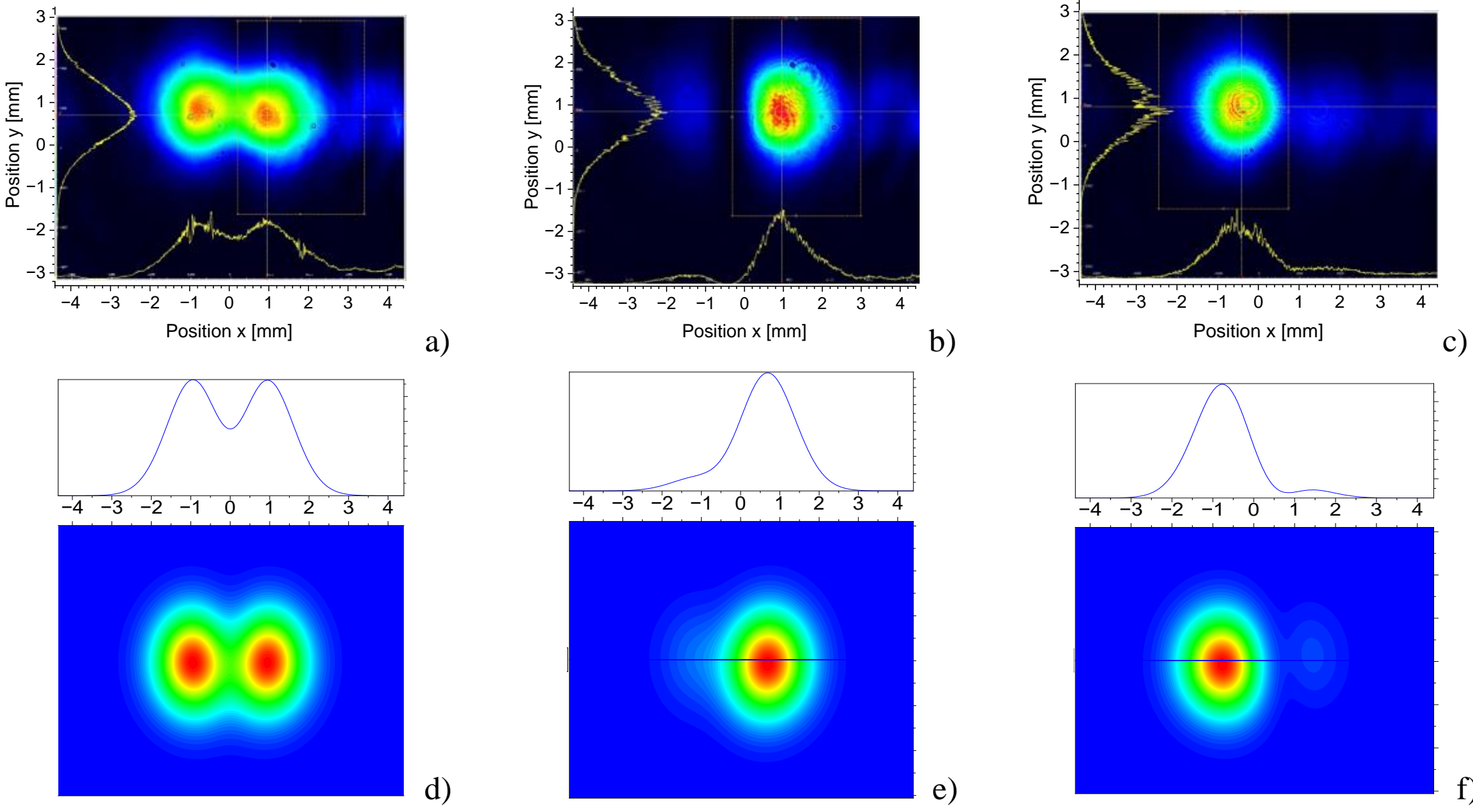

**Fig. 6.** Measured (a, b, c) and calculated (d, e, f) intensity profiles of the reflected beams at a distance of $z = 25$ cm from the grating surface. $\theta_i = \theta_{\text{res}} = 33.53°$, $\gamma = 4°$. (a, d) without polarizers, (b, e) after passing the LCP filter, and (c, f) after passing the RCP filter.

The LCP and RCP (left-handed and right-handed circularly polarized) components of the reflected beam are well separated along the $x$ axis, i.e., they are shifted in opposite directions. The left and right spots have circular polarizations with opposite signs of helicity. Photons with a given helicity are shifted in the opposite direction when the preselection angle changes its sign. This demonstrates the possibility of switching the spatial position of reflected beams with opposite helicity signs by slightly changing the polarization angle of the incident beam.

In Fig. 7, the influence of the preselection angle on the position of the reflected beam with circular polarization is shown. When choosing the optimal preselection angle, a high degree of spin separation is achieved. In Figs. 7g and 7h, the dependencies of the position of the intensity peak $x_m$ and the position of the center of gravity <x$^{\pm}$> on the preselection angle are presented. The centroid position of the reflected beam is given by Eq (8).

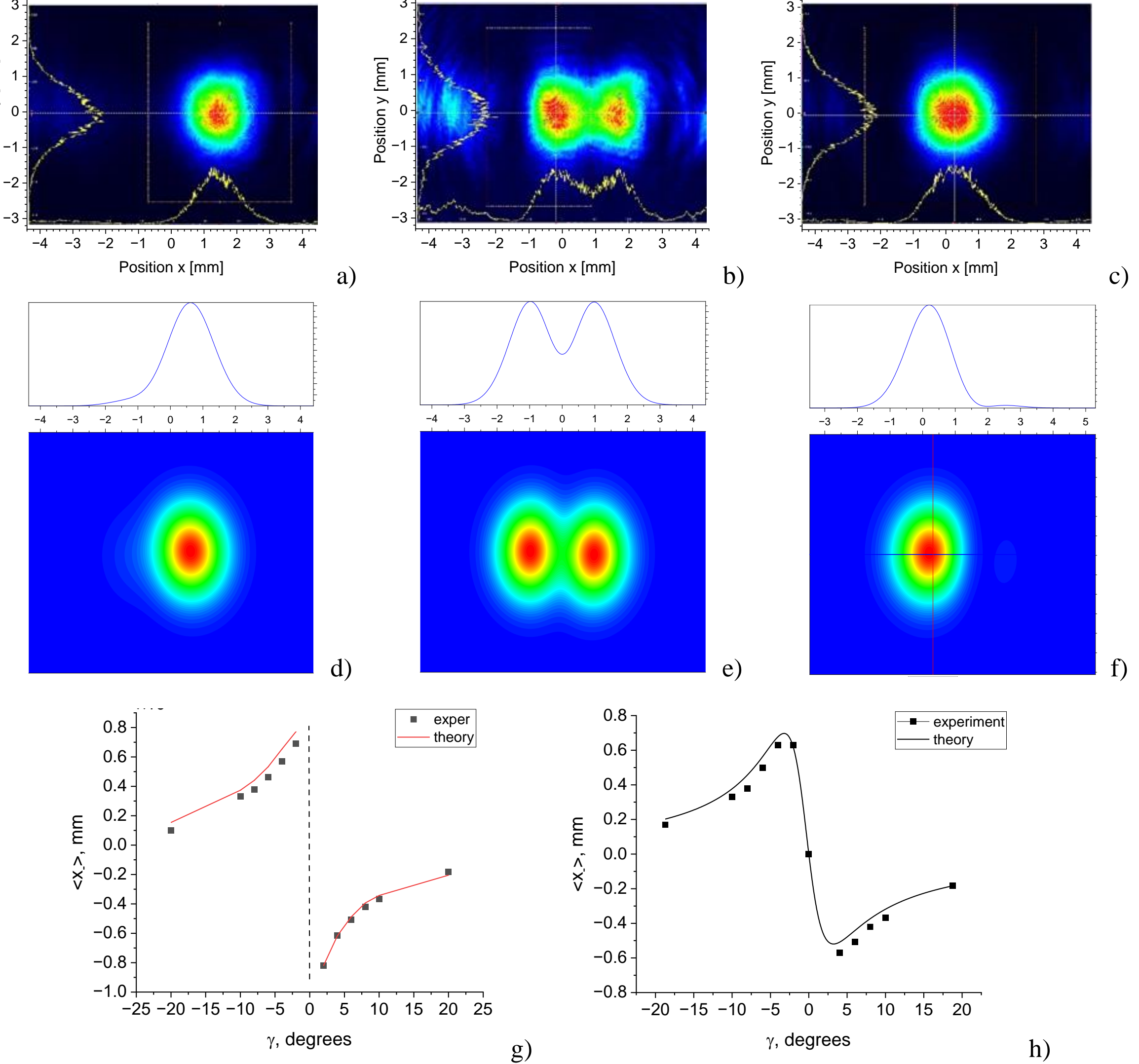

**Fig. 7.** Measured (a, b, c) and calculated (d, e, f) intensity profiles of reflected beams passing through the LCP filter at a distance of $z$ = 25 cm from the grating surface. $\theta_i = \theta_{res}$ = 33.53°, (a, d) $\gamma$ = -4°, (b, e) $\gamma$ = 0°, and (c, f) $\gamma$ = 4°. Angular shifts of the peak intensity position (g) and the centroid position (h) of the reflected beam as function of the preselection angle $\gamma$.

It can be seen, that there is no PSHE at the preselection angle $\gamma$ = 0° (Figs. 7b and 7e). However, the apparent spin splitting occurs at a nonzero preselection angle reaching the maximum separation distance between beams with opposite circular polarization at $\gamma$ = ±4° (Figs. 7a and 7c). The measurements show that the positions of the intensity peaks and centroids are in good agreement with the simulation results.

The final effect of spatial shift amplification is a combination of weak value and propagation amplification. There is an optimal preselection angle (polarization state) for observing the enhanced PSHE (Fig. 7h). Note that a similar effect was found earlier for the prism interface in [42]. It follows from the measurements and theory that the overall weak measurement amplification factor is equal to $A = z/(|\gamma| z_R)$ = 1404.

It follows from the measurements that reflected light under SPR conditions is depolarized and polarization states are mixed because of the interaction of incident light with surface plasmon-polariton waves.

The degree of polarization (DOP) is defined as

$$P = (\eta_1^2 + \eta_2^2 + \eta_3^2)^{1/2}, \qquad (9)$$

where $\eta_1 = \frac{I(0)-I(90)}{I(0)+I(90)}$, $\eta_2 = \frac{I(45)-I(135)}{I(45)+I(135)}$, $\eta_3 = \frac{I_+ - I_-}{I_+ + I_-}$ are the Stokes parameters determined by integral intensity values over the cross section of the reflected beam, $I_+$ and $I_-$ are the intensities of the right- and left-handed circularly polarized beams, respectively.

Note that the degree of linear polarization and the overall degree of polarization decrease significantly with surface plasmon resonance. When the angle of incidence is outside the SPR region, the polarization state of the reflected beam coincides with the polarization state of the incident beam, and the degree of polarization becomes close to unity (Table 1).

**Table 1. Stokes parameters and overall degree of polarization**

| $\delta\theta = \theta_i - \theta_{res}$ | $\eta_1$ | $\eta_2$ | $\eta_3$ | $P$ |
|---|---|---|---|---|
| 0° | 0.537 | 0.028 | 0.379 | 0.657 |
| 0.1° | 0.563 | 0.308 | 0.357 | 0.734 |
| 0.5° | 0.832 | 0.477 | 0.20 | 0.979 |

To quantify the spin separation effect, it is advisable to introduce quantitative characteristics such as spin separation purity (SSP) and/or the degree of circular polarization. Degree of circular polarization $P_c = (I_+ - I_-)/(I_+ + I_-)$ of each spot determines the spin-dependent splitting of the reflected beam. The degree of circular polarization varies depending on the preselection angle and reaches its maximum value at the optimal preselection angle. At a preselection angle of $\gamma = 4°$, the intensity of light with left-handed polarization in the right spot is 20 times higher than the intensity of light with right-handed polarization. This corresponds to a degree of circular polarization exceeding 90%. The spin separation purity can be determined by the parameter $\mu_\pm = \left|I_\pm^L - I_\pm^R\right|/(I_\pm^L + I_\pm^R)$, where the upper indices $L$ and $R$ correspond to the left and right spots, respectively. For the measured powers of 1.32 mW in the right spot and 65 nW in the left, the spin separation purity $\mu_- > 0.9$.

The degree of separation of circular polarization components (spin separation purity) is almost zero at the preselection angle $\gamma = 0°$. This indicates the absence of a spin-dependent shift between the left- and right- circularly polarization components. However, if a non-zero preselection angle is selected, the PSHE can be clearly observed. At the optimal polarization angle of the incident beam (preselection angle), the maximum DOP and the maximum degree of spin separation are achieved. Outside the resonance angle region, the reflected light is predominantly *p*- polarized.

Usually, the SHE of light is associated with the spatial separation of the field components with left- and right-handed circular polarization. Here, along with the well-known PSHE, we demonstrate a new phenomenon of spatial separation of two orthogonal linearly polarized components corresponding to diagonal polarizations. Note that linear polarization is a coherent superposition of two orthogonal circular polarizations, and circular polarization is a coherent superposition of two orthogonal linear polarizations.

The center of gravity (the *x* position average) of reflected beams with diagonal polarization is expressed as

$$\langle x_{\pm 45} \rangle = \frac{\left\langle E^{\pm 45} \middle| x \middle| E^{\pm 45} \right\rangle}{\left\langle E^{\pm 45} \middle| E^{\pm 45} \right\rangle} = \frac{A_\pm}{B_\pm + C}, \tag{10}$$

where $E^{\pm 45}$ are the electric fields, which are linearly polarized along the diagonal directions $+45°$ and $-45°$ to the *x* axis, accordingly,

$A_\pm = -2|z|\cos^2\gamma\left[\left(r_p' \dot{r}_p' + r_p'' \dot{r}_p''\right)\sin\gamma_1 + \left(r_p'' \dot{r}_p' - r_p' \dot{r}_p''\right)\cos\gamma_1\right] \mp |z|\sin(2\gamma)\left[\left(r_s' \dot{r}_p' + r_s'' \dot{r}_p''\right)\sin\gamma_1 + \left(r_s'' \dot{r}_p' - r_s' \dot{r}_p''\right)\cos\gamma_1\right]$;

$B_\pm = 2k_0 z_R \left[\left|r_p\right|^2\cos^2\gamma + |r_s|^2\sin^2\gamma \pm \left(r_p' r_s' + r_p'' r_s''\right)\sin(2\gamma)\right]$; $\quad C = \left|\dot{r}_p\right|^2\cos^2\gamma$; $\quad |z| = \sqrt{z_R^2 + z_r^2}$, $r_{p,s}' = \mathrm{Re}(r_{p,s})$, $r_{p,s}'' = \mathrm{Im}(r_{p,s})$.

In Fig. 8, the measured and calculated intensity profiles of reflected beams are presented at a distance $z$ = 25 cm from the grating surface. Intensity profiles were obtained after passing linear polarizers at an angle of 45° and 135° relative to the direction of polarization of the incident light. It follows from the results of modeling and experiments that spin-independent separation occurs clearly. This indicates that spatial switching of the reflected beam can be implemented for two orthogonal field components corresponding to diagonal linear polarizations. Unlike PSHE, here linearly polarized photons with zero SAM are spatially separated by optimal preselection and post-selection at weak measurements. Since spinless separation occurs, this effect is not due to spin-orbit coupling. Consequently, angular splitting of circularly polarized photons, as well as linearly polarized photons, occurs near the SPR. As can be seen from Fig. 8, the angular shifts are sensitively dependent on both the angle of incidence and the angle of preselection. A particularly high sensitivity is observed with respect to the angle of incidence of the beam (Fig. 8j). Near the SPR, a change in the angle of incidence by only 0.01° causes a spatial displacement of 50 microns at a distance of 25 cm from the grating surface. When the preselection angle $\gamma$ = 0°, there is no spatial separation between the diagonal components.

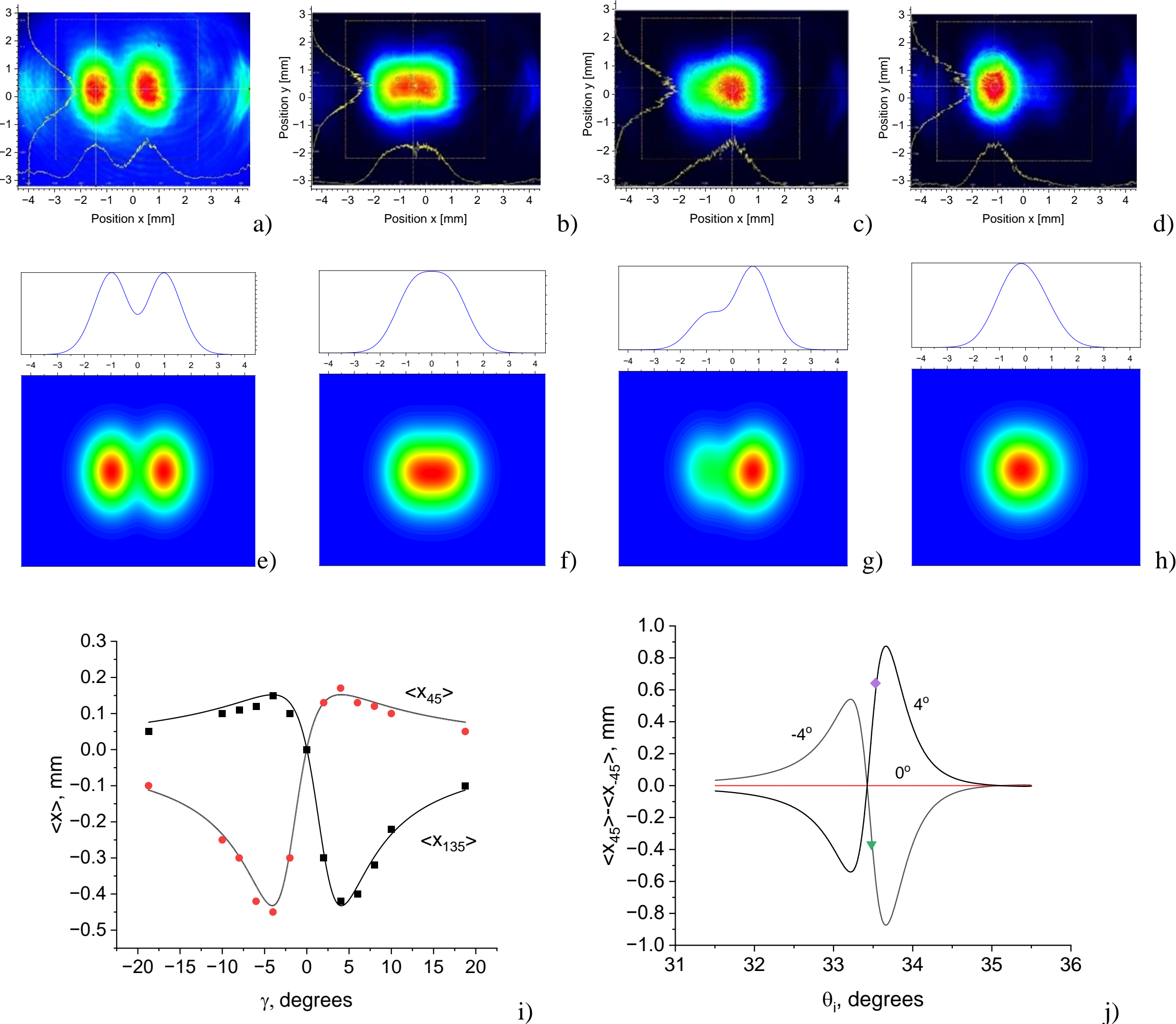

**Fig. 8.** Measured (a-d) and calculated (e-h) intensity profiles of reflected beams at a distance $z$ =25 cm from the grating surface and angle of incidence $\theta_i = \theta_{res} = 33.53°$. (a, e) $\gamma = 0°$, (b, f) $\gamma = 4°$, (c, g) $I(45°)$, (d, h) $I(135°)$, (i) calculated and measured positions of the centers of gravity depending on the preselection polarization angle, and (j) calculated and measured differences between the centers of gravity of reflected beams polarized at an angle of 45° and 135° degrees depending on angle the incidence.

Physically, the observed effect is due to a change in the phase distribution of the field over the cross section of the reflected beam caused by the interaction of incident radiation with a surface plasmon wave. Indeed, the simulations show that the phases of the reflected fields of $p$-polarized component at neighboring spots differ by a value of p, i.e., the electric fields fluctuate in opposite phases. The phase of the $s$-component field remains almost constant over the entire cross section of the reflected beam. Therefore, the left and right spots will be polarized in opposite diagonal directions.

Thus, spin-dependent and spin-independent angular shifts enhanced by SPR have been theoretically and experimentally demonstrated. In contrast to the conventional transverse spin splitting, which is an analogue of the IF shift, here we observed longitudinal spin splitting, which is significantly superior to transverse spin splitting. It is believed that PSHE consists in the separation of photons with opposite spin angular momentum, i.e. in the separation of two orthogonal circular polarizations. Surprisingly, a similar effect can be clearly observed for orthogonally linearly polarized photons. If spin-dependent shifts (spin Hall effect) are the result of spin-orbital interaction, then spin-independent shifts arise because of the phase redistribution of the reflected field at SPR.

Measurements show that the spin splitting increases with a decrease in the size of the focused spot $w_0$ of the incident beam. We experimentally achieved a spin-dependent angular shift of 13.2 mrad between two reflected spots with opposite spin angular momentum when $w_0$ = 17.2 $\mu$m. This displacement corresponds to the spatial separation between the centroids of the reflected spots by 3.1 mm at a distance of 25 cm. These values noticeably exceed the spatial separation of spots with opposite circular polarization observed at the Brewster angle [17-19]. The high sensitivity of the PSHE to the angle of incidence of the light beam near the surface plasmon resonance was demonstrated by choosing the optimal grating parameters (period and depth, dielectric constant of the material) and the incident beam (wavelength, polarization angle). This indicates that a tunable PSHE can be observed by changing the angle of incidence and a preselected polarization angle near the surface plasmon resonance in subwavelength gratings. A sensitivity value of PSHE to changes in the polarization angle of the order of 230 $\mu$m/deg was obtained, which is more than 3 times higher than the values observed when using a planar interface [18]. In addition, the PSHE is very sensitive to changes in the parameters of the grating material, which ensures high accuracy of measurements of small changes in the refractive index. This property can be used in the development of a highly sensitive temperature sensor based on PSHE [43].

The spin separation purity and the degree of circular polarization of each spot reach their maximum value at the optimal preselection angle and the optimal angle of incidence. The polarimetric measurements have shown that the power of the reflected beam at SPR is redistributed into different polarization states. In addition, significant depolarization of light caused by SPR occurs near the resonant incidence angle. Note that although SPR enhances PSHE, which is the result of spin-orbit interaction, at the same time it also leads to the depolarization of light. The fact is that the spin-orbit interaction generates effective birefringence in the medium, which, in turn, is one of the sources of depolarization [44-47].

## 4. Conclusions

In summary, we demonstrated an enhanced photonic spin Hall effect in a SPR subwavelength structure using standard PSHE theory and polarimetric and optimized weak measurement methods. It is shown that changing the angles of incidence and the polarization state are effective ways to control spin angular splitting, especially to increase its magnitude or switch the spatial position of the reflected beam. Extremely large angular shifts have been demonstrated for a nickel subwavelength grating. We found that these large displacements are very sensitive to the angle of incidence and the state of polarization. Polarimetric measurements demonstrate the depolarization of light caused by SPR and the mixing of polarization states. A high degree of spin separation and a degree of circular polarization are achieved with the optimal polarization state and the resonant incidence angle. A new phenomenon of spinless splitting of two orthogonal field components corresponding to diagonal linear polarizations is demonstrated.

The results obtained can be useful in precision measurement and sensing, spectroscopy, spin-selective nanophotonic systems [48, 49], in the development of nanophotonic devices, such as beam splitters and switches, and optical sensors.